# Generalized kinetics of overall phase transition useful for crystallization when assuming heat inertia

Isak Avramov[1], Jaroslav Šesták[2]

[1]Bulgarian Academy of Sciences, BG-1113 Sofia (Bulgaria)

[2]New Technologies - Research Centre of the Westbohemian Region, University of West Bohemia, CZ-301 14  Plzeň (Czech Republic)

Corresponding author: avramov@ipc.bas.bg

**Abstract**

Modeling of process for reaction kinetics is a fashionable subject of publications. The meaning of both the mortality  and fertility terms are mathematically analyzed in details involving variation of their power exponents.

We developed an analogue of the KJMA equation under non-isothermal conditions $\alpha(T) = 1 - \exp\left(-\left(\frac{T}{\theta}\right)^N\right)$ that describes the dependence of degree of transformation α(T) at a constant rate ,$q$, of heating with characteristic temperature $\theta(q)$ and power $N$. This equation is valid even when the activation energy of the process is not constant. We demonstrate that reliable information about the activation energy is obtained when the experimental data are plotted in the following coordinates: log q (heating rate) against log $T_p$ (Peak Temperature).

**Introduction**

Non-isothermal kinetics has become very popular in determining so called reaction mechanism [1-5] and the associated activation energies [6-8] which is a part of famous Arrhenius exponential. These studies are based on well-derived mathematical models [9-11], which however do not have to fully correspond to the actual mechanism of the process described [4], because many of the models used overlap each other [5,11]. Therefore, it is also necessary to look at the other side of the description options that works with generalized equations [5] without the necessary correlation between geometric alignments of individual models [1-5,9-11].

Any phase transition is a complex process involving simultaneous and/or consequent nucleation and growth of separate clusters. Remarkably, there are many processes the time dependencies of which can be described by similar curves [12]. Many of them are not in the field of materials science. Some examples are the pestilence spread among individuals; the propagation of population of animals, or of plants, or of microorganisms, and many others. Most of the existing "epidemic" models make simplifying assumptions about the mechanism of the process. Although the role of disturbance was discussed in details, [13-17], some problems regarding the dynamics of the process remain unsolved. Despite the fact that infection dissemination is essentially a spatiotemporal phenomenon, most models either disregard spatial structure or consider local transmission only [3]. In 1845 while studying the population growth, Verhulst introduced the *logistic functions* [18,19].The initial stage of growth is approximately exponential; then, as saturation begins, the growth slows, and at maturity, growth stops. Actually, all mentioned processes can be described by variants of logistic functions [18-20] that describe the evolution in an environment with a bordered upper limit of the population

**Generalized equation of transformation kinetics**

The idea of logistics functions was first formulated by Verhulst [18,19]. These performs are sigmoid (S-shaped) functions describing the development of fraction α(t) of a new state in a media being subject to exhaustion. It is naturally anticipated [21,22] to assume that the transformation rate is a product of two functions, one of them representing the so called reaction mechanism, *f(α)*, being explicitly dependent on α, only. The other k (t) is usually assumed to depend on time by temperature [21]), which is herewith assumed dependent on time straight away yielding the rate equation

$$\frac{d\alpha(t)}{dt} = k(t)f(\alpha) \qquad (1)$$

This form enables to separate variables having the following solution

$$\int_0^\alpha \frac{d\alpha'}{f(\alpha')} \equiv F(\alpha) = K(t) \equiv \int_0^t k(t')dt' \qquad (2)$$

where *k(t)* is the time derivative of *K(t)*. Even if it sounds strange, the advantage of the equality of the integral functions *F(α) = K(t)* is that we can make conclusions on the nature of one of the integrands $f(\alpha)$ and/or *k(t)* without knowing the explicit form of the other one. To explain this we give examples from kinetics of phase transition, although examples in the rest of the already mentioned processes are also possible. If in a series of experiments we change conditions in such a way that *k(t)* changes but *f(α)* is not affected (for instance we vary the heating rate), the time $t_\alpha$ for which a given value of α is reached, i.e. *F(α)=const,* will give us important information because the condition $\int_0^{t_\alpha} k(t')dt' = const$ is also required. In the following we use the reciprocal of *k*, the characteristic time $\tau = 1/k$, because this is more suitable as the dimension of $\tau$ is [time]. If $\tau$ does not change with time (which is possible only under isothermal conditions) we have the simplest solution

$$K(t) = \frac{t}{\tau} \tag{3}$$

The simplest assumption, is that *f(α)* is proportional to the product of *α* and *(1-α)*.

$$f(\alpha) = \alpha(1-\alpha) \tag{4}$$

The term α is an assumption that the process proceeds as a first order autocatalytic reaction (Prout and Tompkins [23]) while the term *(1-α)* accounts that the process can proceed only in that part of the system that is not yet transformed. In the other worlds [22] it is called *mortality*,α, which subsists the reactant disappearance and the product formation and the complementary *fertility* of a reactant yet ready to act in response*(1-α)*. There are several assumptions on *f(α)*, the corresponding *F(α)* functions /24,25/ and the solutions for the time dependence of the degree of transformation α. A generalized assumption [19,20] is that the transformation proceeds as an autocatalytic reaction of order *p*.

$$\frac{d\alpha}{dT} = \left[\frac{1}{(1-p)\tau}\right]\alpha^p(1-\alpha) \tag{5}$$

As far as the explicit form of characteristic time is not specified for the moment, here for convenience we introduce the numerator *(1-p)*.

If there are enough reasons to believe that the conditions are not essentially changing during the process, i.e., if $\tau$ can be assumed constant, it is appropriate to replace in the table *K(t)* by $t/\tau$ (later on we shall discuss also some possible $\tau(t)$ – dependencies). For instance, in the case of kinetics of phase transitions, $\tau(t)$ depends on time if the temperature is changing.

Several solutions of Eq(5) for different values of the order parameter *p* are discussed in / 25,26. Here we will try to make a universal solution good for any *p* value. We will make this solution a variant of the known KJMA equation, which has the form

$$\alpha = 1-\exp\left(-\left(\frac{t}{\tau}\right)^n\right) \qquad (6)$$

Here it will be demonstrated that the KJMA equation follows if transformation proceeds as an autocatalytic reaction of order $p$ (Eq(5)). It is convenient to introduce the parameter $n = \frac{1}{1-p}$. In this way Eq(5) becomes

$$\frac{d\alpha}{dT} = \left[\frac{n}{\tau}\right]\alpha^{\frac{n-1}{n}}(1-\alpha) \qquad (7)$$

Since α is small parameter we use the approximation

$$\ln x \approx x - 1 , \qquad (8)$$

only this time in opposite direction, i.e. we introduce a logarithmic term by substituting x - 1. In particular, here we use it in the form

$$\alpha^{\frac{n-1}{n}} \approx -\left[\ln(1-\alpha)\right]^{\frac{n-1}{n}} \qquad (9)$$

When Eq(9) is introduced into Eq(7), a differential equation is obtained, the solution of which leads to universal validation of KJMA equation(6) with n being the so called Avrami parameter.

**Kinetics of phase transition under non-isothermal conditions**

If τ is not constant, we need to have it explicit form to solve $K(t)$. For instance, if temperature of the system is changing with time, τ is also changing according to expression

$$\tau(T) = \tau_o \exp\left(\frac{E}{RT}\right) \qquad (10)$$

depending on the activation energy $E$. Since at low temperatures $\tau(T)$ is very large, no process is running there and we can assume that the heating at constant rate $q$ has started at $T=0$ [K] and $t = \frac{T}{q}$, so that

$$\frac{d\alpha}{f(\alpha)} = \frac{dT}{q\tau(T)} \qquad (11)$$

The rate parameter *K(t)* is becoming

$$K(T) = \int_0^T \frac{e^{-\frac{E}{RT'}}}{q\tau_o} dT' \qquad (12)$$

The integral of Eq(12) is solved only after it is assumed that the activating energy E is a constant. This seriously limits the scope of its possible applications. So we have to find a way to replace the exponential term in expressing the time of crystallization τ (Eq(10)), without changing the nature of the assumption of activation energy. Although the crystallization interval depends on the rate of heating, the process proceeds in a relatively narrow temperature range. For this reason, we involve a reference temperature $T_r$ somewhere in the middle of the interval. (Later we demonstrate that $T_r$ exact position is not significant) Thus, a non-dimensional temperature $\frac{T_r}{T}$, which varies around a unit with a deviation of no more than 5%, is introduced. Using this temperature we can express the activation energy and develop it in power series in the following form:

$$\frac{E}{RT} = \frac{E}{RT_r}\left(\frac{T_r}{T}\right) \approx \frac{E_r}{RT_r} + \left(\frac{d\left[\frac{E}{RT_r}\left(\frac{T_r}{T}\right)\right]}{d\left(\frac{T_r}{T}\right)}\right)_{\frac{T_r}{T}=1} \left(\left(\frac{T_r}{T}\right)-1\right) \qquad (13)$$

The term in square brackets is the apparent activation energy $\mathcal{E}$

$$\mathcal{E} \equiv \left(\frac{d\left[\frac{E}{RT_r}\left(\frac{T_r}{T}\right)\right]}{d\left(\frac{T_r}{T}\right)}\right)_{\frac{T_r}{T}=1} = \frac{E_r}{RT_r} + \left[\frac{d\left[\frac{E}{RT_r}\right]}{d\left(\frac{T_r}{T}\right)}\right]_{\frac{T_r}{T}=1} \qquad (14)$$

For constant activation energy this is just its dimensionless form $\mathcal{E} = \frac{E_r}{RT_r}$. It is useful to compare the term in square brackets on the right side of Eq(14) with the dimensionless activation energy in the form

$$\left[\frac{d\left[\frac{E}{RT_r}\right]}{d\left(\frac{T_r}{T}\right)}\right]_{\frac{T_r}{T}=1} = (a-1)\frac{E_r}{RT_r} \qquad (15)$$

Here *a* plays a role of "fragility" parameter and accounts to what extent the activation energy depends on temperature. For *a=1* activation energy is constant. With this fragility the apparent activation energy $\mathcal{E}$ becomes

$$\mathcal{E} = a\frac{E_r}{RT_r} \qquad (16)$$

Note that the apparent dimensionless activation energy $\mathcal{E}$ has relatively large value $\mathcal{E} \geq \frac{E_r}{RT_r}$.

For the variable $\frac{T_r}{T}$ the approximation given by Eq(8) is valid, so that

$$\frac{E}{RT} = \frac{E_r}{RT_r} + \mathcal{E}\left(\left(\frac{T_r}{T}\right) - 1\right) = \frac{E_r}{RT_r} + \mathcal{E}\ln\left(\frac{T_r}{T}\right) \qquad (17)$$

In this way, the crystallization time (Eq(10)) is released from the exponential term because

$\tau = \tau_o \exp\left[\frac{E_r}{RT_r}\right]\exp\left(\ln\left(\frac{T_r}{T}\right)^{\mathcal{E}}\right)$ easily transforms to

$$\tau(T) = \tau_r\left(\frac{T_r}{T}\right)^{\mathcal{E}} \qquad (18)$$

Here $\tau_r$ is the crystallization time at the reference temperature. Note that $\tau(T)$ is valid for every form of apparent activation energy $\mathcal{E}$, no matter how it depends on the temperature. This time the temperature dependence of degree of transformation $\alpha(T)$ is solved easily

$$\alpha(T) = 1 - \exp\left(-\left(\frac{T_r}{q\tau_r(\mathcal{E}+1)}\left(\frac{T}{T_r}\right)^{\mathcal{E}+1}\right)^n\right) \qquad (19)$$

In order to make the expression appear more friendly, we introduce two new variables: the parameter

$$N = n(\mathcal{E}+1) \qquad (20)$$

and the characteristic temperature

$$\theta = \left[q\tau_r T_r^{\mathcal{E}}(\mathcal{E}+1)\right]^{\frac{1}{\mathcal{E}+1}} \qquad (21)$$

. With these notations we arrive to the main expression presenting the temperature dependence of degree of transformation $\alpha(T)$

$$\alpha(T) = 1 - \exp\left(-\left(\frac{T}{\theta}\right)^N\right) \qquad (22)$$

Here $\theta$ is the characteristic temperature at which the transformation degree $\alpha(\theta)=1-1/e$. Moreover, it can be shown the peak temperature $T_p$ of the $\alpha(T)$ curve is approximately

$$T_p \approx \theta \qquad (23)$$

By combining Eq(21) and Eq(23), it is seen that when the experimental data are set to logarithmic coordinates (log $q$ vs log $T_p$), a straight line is expected from the slope of which $\mathcal{E}$ can be directly determined

$$\log q = (\mathcal{E}+1)\log T_p - \mathcal{E}\log T_r - \log(\mathcal{E}+1) - \log(\tau_r) \qquad (24)$$

Moreover, the intercept, Y, gives information about the crystallization time $\tau_r$ at reference temperature.

$$Y = -\left(\mathcal{E}\log T_r + \log(\mathcal{E}+1) + \log(\tau_r)\right) \qquad (25)$$

To evaluate the value of $\tau_r$, we use the $\mathcal{E}$ value already obtained from the slope and we take for $T_r$ the mean value of the temperature range in which the study was conducted. Obviously, the accuracy of the determination is a function of the width of this temperature range

**Note on a still forgotten effect of heat transfer and dispersion**

Presenting our mathematical analysis deliberately ignores the practical effects of real measurements, which specifically for the dynamic mode of thermal analysis include energy transfer, i.e., heat transfer and dissipation in the sample body [26-29]. Every amount of externally delivered or internally produced heat in the sample must be equilibrated within the sample´s thermal capacity, $C_p$, that is by adjusting the molecular motions of all the species present there. All motions notably have the same inertia behavior as any macroscopic motion described in the Newton´s historical laws. Let us remember that heat inertia is physically the product of heat capacity, $C_p$, and the rate of its temperature changes, $\Delta T/dt$. This is well comparable with its mechanical equivalent: the mechanical momentum, which is the product of mass and its movement - velocity. Thus $C_p$, plays the role of mass and $\Delta T/dt$ its speed. While no one would question the validity of mechanical momentum, heat inertia resulting from the time-honored Newton cooling law continues to be overlooked. Some kineticists are reasoning that its incorporation would be too complicated and would impair the computational throughput. Others say that if thermal inertia has not been applied in kinetics over the past 300 years, there is no reason to apply it now and in the future. Still others use philosophical resignation, saying that science is never accurate and it only approximates the truth - as an excuse for knowingly using inaccurate models while the ways to correct them has been known for a long time. This phenomenon has been detailed in our recently published papers [26-29] on which we refer the readers´ kind attention and therefore we have not deliberately included this phenomenon in our deeper analysis.

**Conclusion**

The existing framework of modeling reaction mechanism, involving effortless interpretation of many yet unclear features of heterogeneous kinetics (such as the physical meaning of the Arrhenius constants) became unwearyingly trickery [6 - 8]. As a common consequence the overall use of the activation model has apparently led to a long-term stagnant study of mechanisms of solid-state reactions. Interpretation of experimental results of differential thermal measurements (DTA) has become inconclusive [26], which, however, is not the subject of our communication. We have successfully solved the problem of an arbitrary form of activating energy and the order of the autocatalytic reaction. In this way, we were able to offer coordinates to interpret experimental results in order to extract reliable information.

**Acknowledgement:**

The support of the Project "Oberflächeninitiierte Mikrostrukturbildung in Glaskeramiken" (Project number: 382920141 DGF-project ) is appreciated

The present work was supported by (the CENTEM Project, Reg. No. CZ.1.05/2.1.00/03.0088 that is co-funded from the ERDF as a part of the MEYS—Ministry of Education, Youth and Sports OP RDI Program and, in the follow-up sustainability stage supported through the CENTEM PLUS LO 1402).